\documentstyle[12pt,aps,epsfig]{revtex}

\def\laq{\raise 0.4ex\hbox{$<$}\kern -0.8em\lower 0.62
ex\hbox{$\sim$}}
\def\gaq{\raise 0.4ex\hbox{$>$}\kern -0.7em\lower 0.62
ex\hbox{$\sim$}}

\def\NPB{{\em Nucl. Phys.} B}
\def\PLB{{\em Phys. Lett.}  B}
\def\PRL{\em Phys. Rev. Lett.}
\def\PRD{{\em Phys. Rev.} D}

\def\be{\begin{equation}}
\def\ee{\end{equation}}
\def\bea{\begin{eqnarray}}
\def\eea{\end{eqnarray}}

\begin{document}

\preprint{\vbox{\baselineskip=12pt
\rightline{BGU-PH-98/06}
\vskip1truecm}}

\title{Production and Detection of \\
 Cosmic Gravitational Wave Background \\ in String Cosmology \\ \ }

\author{ RAM BRUSTEIN \\ \ }

\address{Department of Physics, Ben-Gurion University, 
Beer-Sheva 84105, Israel \\ \ }

\maketitle{
\begin{abstract}
String cosmology models predict a cosmic background of 
gravitational waves produced during a period of dilaton-driven inflation. I
describe the background, present astrophysical and cosmological bounds on it,
and discuss in some detail how it may be possible to detect it with large 
operating and planned gravitational wave detectors.  The possible use of 
smaller detectors is outlined.
  \end{abstract}

\section{Introduction}

A robust prediction of models of string cosmology which realize the
pre-big-bang scenario \cite{sfd,pbb} is that our
present-day universe contains a cosmic gravitational wave background 
\cite{relic,gg,gwg},  with a spectrum which is quite different than that
predicted by  other early-universe  cosmological
models \cite{grishchuk,turner,allen,maggiore}.  In the pre-big-bang scenario the
evolution of the universe starts from a state of very small curvature and
coupling and then undergoes a long phase of dilaton-driven kinetic inflation
reaching nearly Planckian energy densities \cite{exit1},  and at some later time
joins smoothly standard  radiation dominated cosmological evolution, thus giving
rise to a singularity free inflationary cosmology. 

In this paper I describe  the cosmic 
gravitational  wave background predicted by models of string cosmology 
and present numerical
estimates for spectral parameters,
I review astrophysical  and cosmological bounds  on the spectrum's shape 
and strength, and
 show that currently operating and planned large gravitational  wave detectors
could further constrain the spectrum and perhaps even detect it. I discuss
detection strategies and compare the efficiency of different types of detectors.
Finally, I outline the possible use of small resonators and the use of the
``memory effect" to detect the background or constrain its parameters.

Because the gravitational interaction is so weak, a 
background of gravitational radiation decouples from  matter in the
universe  at very early times and carries with it information on the state of the
universe  when energy densities and temperatures were extreme.
The weakness of the gravitational interaction makes a detection of such a
background very hard, and necessitates a strong signal. String
cosmology provides perhaps the strongest source possible: the whole universe, 
accelerated to nearly Planckian energy densities. Although in this paper I
use particular string cosmology models, the main conclusions will remain valid
for all models in which the universe spends a finite time at near
Planckian energy densities. 

A discovery of any primordial gravitational  wave background, and in particular,
the one predicted by string cosmology,  could provide unrivaled
exciting information on the very early universe. Such a discovery
will confirm the basic principles used to theoretically derive the background. 
For example, it will confirm the validity of quantum mechanics as we know it 
all the way up to the Planck scale.

\section{Cosmic gravitational wave background in string cosmology}
\label{section:first}

In models of string cosmology \cite{relic} (see also \cite{sfd,pbb}), 
the universe passes through
two early inflationary stages.  The first of these is called the
``dilaton-driven" period and the second is the ``string" phase.  Each
of these stages produces stochastic gravitational radiation by  the standard
mechanism of amplification of quantum fluctuations \cite{mukh}. 
Deviations from homogeneity and isotropy of the metric field are generated by 
quantum  fluctuations around the homogeneous and isotropic background, and then
amplified by the accelerated expansion of the universe. The transverse and 
traceless part of these fluctuations are the gravitons. In practice, we
compute graviton production by solving  linearized
perturbation equations with vacuum fluctuations boundary conditions. 
The production strength of gravitons depends on the curvature and coupling. Since
at the end of the accelerated expansion phase curvatures reach the 
string curvature, and the coupling reaches approximately the present coupling,
graviton production is expected to be at the strongest possible level.

In order to describe the background of gravitational radiation, it is
conventional to use a spectral function 
$
\Omega_{\rm GW}(f) = {1 \over \rho_{\rm critical}} {d \rho_{\rm
GW} \over d \ln f},
$ 
where $ d \rho_{\rm GW}$ is todays energy density in
stochastic gravitational waves (GW) in the frequency range $ d \ln f$, and
$\rho_{\rm critical}$ is the critical energy-density required to just
close the universe,
$
\rho_{\rm critical} = { 3 c^2 H_0^2 \over 8 \pi G} \approx 1.6 \times
10^{-8} {\rm h}_{100}^2 \rm \> ergs/cm^3.
$
The Hubble expansion rate $H_0$ is the rate at which our universe
is currently expanding,
$
H_0 = {\rm h}_{100} \> 100 \> {\rm  Km \over sec-Mpc} = 
3.2 \times 10^{-18}{\rm h}_{100} Hz.
$
 ${\rm h}_{100}$  is believed to lie in the
range $0.5 < {\rm h}_{100} < 0.8$. The spectral function is related to the
dimensionless strain $h$, 
$\Omega_{\rm GW}(f) \simeq 10^{36} h_{100}^{-2}(f/ {\rm Hz})^2 h(f)^2$ 
and to the strain in units $1/\sqrt{Hz}$, $\sqrt{S_h(f)}$, 
\begin{equation}
\Omega_{\rm GW}(f) = 1.25 \times 10^{36} h_{100}^{-2} (f/ {\rm Hz})^3 S_h(f).
\end{equation}
The spectrum of gravitational radiation produced during the dilaton-driven
(and string) phase was estimated in \cite{relic} 
(see also \cite{gg,bggmv,sduality,bh,bmuv}). It is approximately given by
\begin{equation}
\Omega_{GW}(f) =
z_{eq}^{-1} g_s^2
 \left(f \over f_S\right)^3
\left[ 1+ z_S^{-3} \left(g_1\over g_S\right)^{2}\right],  f<f_S,
\end{equation}
where some logarithmic correction factors were dropped.
The coupling $g_1$ is today's
coupling, assumed to be constant from the end of the string phase,
$g_S$ is the coupling at the end of the dilaton-driven phase, and $f_S$ is the
frequency marking the end of the dilaton-driven phase. The frequency $f_1=f_S
z_S$ is the frequency at the end of the string phase, where $z_S$ is the total
red-shift during the string phase and $z_{eq}\sim 10^4$ is the red-shift from 
matter-radiation equality until the present. We will present a more quantitative
and detailed estimate of spectral parameters later.

The spectrum can be expressed in a more
symmetric  form \cite{sduality},    
\begin{equation}
\Omega_{GW}(f)
= z_{eq}^{-1} g_1^2 \left(\frac{f}{f_1}\right)^3  
 \left[ z_S^3 (g_S/g_1)^2+ z_S^{-3} (g_S/g_1)^{-2}\right].
\end{equation}
  Note that the spectrum is
invariant under the exchange  $z_s^3 (g_s/g_1)^2 \leftrightarrow z_s^{-3}
(g_s/g_1)^{-2}$
 and that this implies a lower bound on the  spectrum,  
$
\Omega_{GW}(f) \gaq\ 2 z_{eq}^{-1} g_1^2 \left(\frac{f}{f_1}\right)^3.$
The lower bound is obtained for the ``minimal spectrum" 
with $z_S=1$ and $g_S/g_1=1$ describing a cosmology with almost no intermediate
string phase.

In the simplest model, which we will use to estimate the spectrum and prospects 
for its detection, the spectrum depends upon four parameters.   The first pair
of parameters  are the maximal frequency $f_1$  above which gravitational
radiation is not produced and $g_1$, the coupling at the end of the string
phase. The second pair of these are $z_S$ and $g_S$. The second pair of
parameters can be traded for  the  frequency ${f_{S}}=f_1/z_S$ and the
fractional energy density $\Omega^{\rm S}_{\rm GW}=\Omega_{\rm GW}(f_S)$  
produced at the end of
the dilaton-driven phase. At the moment, we cannot compute $g_S$ and $z_S$ from
first principles, because they involve knowledge of the evolution during the
high curvature string phase. We do, however, expect $z_S$  to be quite large.
Recall that $z_S$ is the total red-shift during the string phase, 
 and that during this phase 
the curvature and expansion rate are  approximately string scale,
therefore, $z_S$ grows roughly exponentially with the duration (in string times)
of this phase. Some particular exit models \cite{exit2} suggest
that $z_S$ could indeed be quite large. I
cannot estimate, at the moment, a likely range for the ratio $g_1/g_S$  except
for the reasonable assumption $g_1/g_S>1$. We prefer to concentrate on the
features of the spectrum that can be computed theoretically as cleanly as
possible. Since at the moment the best understood part of the spectrum is the
part produced during the dilaton-driven phase, we concentrate our attention on
parameters associated with this phase.

A useful approximate form for the spectrum in the range $z_S>1$ and
$g_1/g_S\gaq 1$  is the following~\cite{rb}
\be \label{e:approx}
\Omega_{\rm GW}(f)=
\cases{
{\Omega^{\rm S}_{\rm GW} (f/{f_{\rm S}})^3}  &  { $f<{f_{\rm S}}$} \cr
 & \cr
{\Omega^{\rm S}_{\rm GW} (f/{f_{\rm S}})^{\beta} } &   ${f_{\rm
S}}<f<f_1$ \cr
 & \cr
{0} &   $f> f_1.$}
\ee
where
$
\beta=\frac{\log\left[\Omega_{\rm GW}(f_1)/
\Omega^{\rm S}_{\rm GW}\right]}{\log\left[f_1/{f_{\rm S}}\right]}
$
is the logarithmic slope of the spectrum produced in the string phase 
(see also other models \cite{spgravitons}). The corresponding
spectral density $S_h$, in units of $Hz^{-1}$, is given by
\be \label{e:sapprox}
S_h(f)=
\cases{
{S^{ S}_{ h} }  &  { $f<{f_{ S}}$} \cr
 & \cr
{S^{ S}_{ h} \ (f/{f_{ S}})^{\beta-3} } &   ${f_{
S}}<f<f_1$ \cr
 & \cr
{0} &   $f>f_1$.}
\ee
Note that $S_h$ is constant during the dilaton-driven phase.  The form
(\ref{e:sapprox}) is particularly useful in comparing sensitivities of different
detectors.

If we assume that there is no late entropy production and make
reasonable choices for the number of effective degrees of freedom, then
two of the four parameters may be determined in terms of the Hubble
parameter $H_*$ at the onset of radiation domination immediately
following the string phase.

We turn now to obtain numerical estimates of $f_{S}$ and 
$\Omega^{\rm S}_{\rm GW}$. Our assumptions are somewhat
different than those used in \cite{peak}, but the resulting range is similar. 
To obtain estimates for the  spectral parameters we must assume some late time
background cosmology. Here we assume  standard cosmology in a flat universe
without a cosmological constant. A different choice of late time background
cosmologies will lead to calculable changes in these estimates. To obtain numerical
estimates for the spectral parameters it is useful to consider the ``minimal
spectrum", in which the the dilaton-driven inflationary phase connects almost
immediately to standard radiation-dominated evolution. For the minimal spectrum
$z_S=1$, $g_S/g_1=1$, $f_{1}=f_S$.

We start our discussion with the frequency axis.
For the minimal spectrum, the frequency of the end-point $f_1$  
today is given by the frequency which just reenters the horizon at the
beginning of the radiation dominated phase,  red-shifted to its present value, 
\begin{equation}
f_{1}=f_{*}/z_{*}=\frac{H_{*}}{2 \pi} \frac{1}{z_{*}}
\label{ftoday}
\end{equation}
where $H_{*}$ is the Hubble parameter at the end of the string phase 
and $z_{*}$ is the red-shift since then, given as the ratio of the
scale factors today and then $z_{*}={a_0}/{a_{*}}$.
We use entropy considerations to evaluate $z_{*}$.  

Let us first assume that entropy is approximately conserved during the evolution  
from the  end of the string phase until today. This must be an approximation.
Entropy cannot be  absolutely conserved because some non-adiabatic processes, 
such as the relaxation of the dilaton and other moduli towards the minimum of
their potential,  are expected.
If entropy is approximately conserved, we obtain 
${\rm g}_s(t_0) T_0^3 a_0^3={\rm g}_s(t_*) T_*^3 a_*^3$ 
from which we may calculate 
\hbox{$z_*=\frac{T_*}{T_0} 
\Biggl[\frac{{\rm g}_s(t_*)}{{\rm g}_s(t_0)}\Biggr]^{1/3}\!\!,$} where 
$\hbox{\rm g}_s=\!\!\sum\limits_{\ i_{bosons}}\!\! \hbox{\rm g}_i (\frac{T_i}{T})^3+
\frac{7}{8}\!\!\sum\limits_{\ i_{fermions}}\!\! \hbox{\rm g}_i (\frac{T_i}{T})^3$ 
measures the effective number of degrees of freedom and should not be confused
with the string coupling parameter at the beginning of the string phase $g_S$.
In the previous equations, and in the rest of the paper, a subscript 0
refers to the present values of various quantities.
Since ${\rm g}_s(t_0)=3.91$ and $T_0=2.74 K$  are known (see, for example,
\cite{eu}), 
$z_*$ is given by 
\begin{equation}
z_*=\frac{T_*}{2.74 K} \Biggl[\frac{{\rm g}_s(t_*)}{3.91}\Biggr]^{1/3}\!\!.
\label{z*1}
\end{equation}

Assuming local thermal equilibrium and radiation domination at $t_*$ 
we may relate $T_*$ to $H_*$ in a standard way 
\begin{equation}
T_*=H_*^{1/2} m_{pl}^{1/2} \left[\frac{90}{8 \pi^3  
{{\rm g}}_\rho}\right]^{1/4}
\label{T*}
\end{equation}
where $G_N\equiv1/m_{pl}^2$ and $\hbox{\rm g}_\rho=\!\!\sum\limits_{\
i_{bosons}}\!\! \hbox{\rm g}_i (\frac{T_i}{T})^4+ \frac{7}{8}\!\!\sum\limits_{\
i_{fermions}}\!\! \hbox{\rm g}_i (\frac{T_i}{T})^4$  . 
Substituting $T_*$ from eq.(\ref{T*}) into eq.(\ref{z*1})
\begin{equation}
z_*=\frac{H_*^{1/2} m_{pl}^{1/2}}{2.74 K} 
{{\rm g}}_\rho^{-1/4} {{\rm g}}_s(t_*)^{1/3}
\left[\frac{90}{8 \pi^3} \right]^{1/4} 3.91^{-1/3},
\label{z*2}
\end{equation}
and substituting $z_*$ from eq.(\ref{z*2}) into eq.(\ref{ftoday}) we obtain
\begin{equation}
f_{1}=1.2\times10^{11} Hz \times\left[
\left(\frac{H_*}{ m_{pl}}\right)^{1/2} 
 {\rm g}_\rho^{1/4} {\rm g}_s(t_*)^{-1/3}\right]. 
\label{ftoday3}
\end{equation}
Since we expect $H_*$ to be less than $m_{pl}$ and of the order 
of the string scale, $M_{s}\sim 5\times 10^{17} GeV$  it is convenient to 
express $H_*$  as 
$\left(\frac{H_*}{ m_{pl}}\right)^{1/2}= 0.20
\left(\frac{H_*}{ 5\times 10^{17} GeV}\right)^{1/2}$.  In addition, since both 
${{\rm g}}_\rho$ and  ${{\rm g}}_s(t_*)$ are expected to be  approximately
equal and much larger than  the standard model values, ${\rm g}_s\sim 100$, 
we assume for simplicity that they are equal, denote their common value 
as ${\rm g}_*$ and parametrize them as
${{\rm g}}_\rho^{1/4} {{\rm g}}_s(t_*)^{-1/3}=0.56\ [{\rm
g}_*/1000]^{-1/12}$.   Putting everything together we obtain
\begin{equation}
f_{1}=1.3\times10^{10} Hz \times\left[
\left(\frac{H_*}{ 5\times 10^{17} GeV}\right)^{1/2} 
 \left(\frac{{\rm g}_*}{ 1000}\right)^{-1/12}\right]. 
\label{ftodayfinal}
\end{equation}
Equation (\ref{ftodayfinal}) is the final result for the 
end-point frequency of the minimal spectrum assuming no entropy production 
and initial 
radiation domination just after the string phase  
(but not necessarily afterwards). 

If the transition from the dilaton-driven phase to radiation domination
is not immediate, as expected, and we neglect the effects of the backreaction of 
the produced particles on the background cosmology, then $f_S$ is 
simply red-shifted by the total amount of 
red-shift accumulated during the string phase, $f_S= f_1/z_S$.

If entropy was not even approximately conserved since the end of the stringy phase,
then according to the second law of thermodynamics, entropy had to increase.
This means that the value of $f_{1}$ in eq.(\ref{ftodayfinal})
 is an  upper bound on $f_{1}$, as shown in \cite{peak}. 
The spirit of our model, in general, favors  approximate entropy conservation.

The analysis of the amplitude axis depends more strongly on the details of
the model, and therefore less accurate. 
Our starting point is the estimate of GW energy density
\cite{relic,sduality,bh,bmuv} 
\begin{equation}
\frac{d \rho_{GW}}{d^3 x d \ln k}= C k_1^4 (g_S/g_1)^2 (k/k_S)^3,
\label{ampep1}
\end{equation}
where $C$ is a numerical coefficient of order 1, depending on the details of
the matching procedure between phases. Deviding by the critical energy density
$\rho_c$ we obtain 
\begin{equation}
\Omega_{GW}(f)=
 C   \frac{(2\pi f_1)^4}{\frac{3}{8\pi}H_0^2 m_{pl}^2} \ (g_S/g_1)^2\
(f/f_S)^3 \label{ampep}
\end{equation}
Substituting $f_1$ from eq.(\ref{ftodayfinal}), 
setting $C$ to unity for the purpose of obtaining some definite answer,  and
using  known numerical values we obtain   
\begin{equation}
\Omega_{GW}^S= 1.3 \times 10^{-7} h_{100}^{-2} 
\left(\frac{{\rm g}_*}{1000}\right)^{-1/3}\ 
\left(\frac{H_*}{ 5\times 10^{17} GeV}\right)^2\ (g_S/g_1)^2
\label{ampep2}
\end{equation}
Equation (\ref{ampep2}) reflects the absolute normalization of the amplitude
provided  by the uncertainty principle, however, it does involve an arbitrary
numerical factor $C$, which was set to unity and which depends on details
of the background evolution.

This completes the 
determination of the end-point coordinates for the minimal spectrum. 

For non-minimal spectra the effect on the position of $\Omega_{GW}^S$ is
more complicated. One important effect is that $g_S$ is no longer equal to $g_1$ 
and could be much smaller. There are some indications that $g_S$
could  be a free parameter, depending on the initial conditions. 

If entropy is not even approximately conserved, namely, if a phase of massive
 entropy production occurs at some later time then 
the amplitude of modes still outside
the horizon does not change while  the amplitude of modes inside
the horizon decreases. 
Details of entropy production are important, for example, if GW are also 
produced by the entropy creation process then their spectrum gets modified, 
some parts are enhanced while other suppressed.

In summary, the estimated range of spectral parameters, if there is no
substantial entropy production is as follows,
$$
f_S= { 1.3 \times 10^{10} } \frac{1}{z_S} \left( { H_{\rm *} \over 5 
\times 10^{17} \> {\rm GeV}}\right)^{1/2} 
\left( { {\rm g}_{*} \over 1000} \right)^{-1/12}\ Hz, 
$$
$$
\Omega_{\rm GW}^S = 1.3 \times 10^{-7} { h}_{100}^{-2} 
\left(\frac{H_*}{ 5\times 10^{17} GeV}\right)^2
  \left( { {\rm g}_{*} \over 1000} \right)^{-1/3} \ (g_S/g_1)^2,
$$
and
$$
S_h^S= 1.0 \times 10^{-43} 
\left(\frac{H_*}{ 5\times 10^{17} GeV}\right)^2
  \left( { {\rm g}_{*} \over 1000} \right)^{-1/3} 
\left( { f_S \over Hz} \right)^{-3}\ (g_S/g_1)^2 \ Hz^{-1}
$$

\vspace{-1in}
\begin{figure}
\begin{center}
\psfig{figure=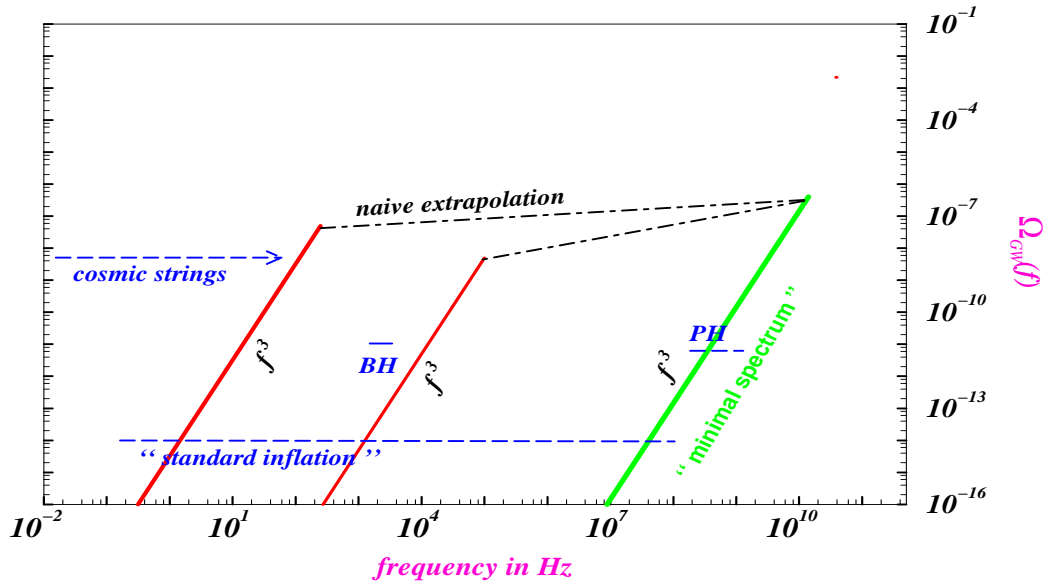,width=5.5in,height=6.5in}
\end{center}
\vspace{-1.8in}
\caption{\small Spectrum of GW background. The minimal spectrum discussed  in
the text and two other possible spectra are shown. Also shown are estimated
spectra in other cosmological models.} 
\label{fig:ogw}
\end{figure}

Entropy production, roughly speaking, lowers $f_S$ and has a more complicated
effect on $\Omega_{\rm GW}^S$. For an example of a possible effect of entropy
production see \cite{peak}.

Spectra for some arbitrarily chosen parameters and possible backgrounds
from other cosmological models are shown in Fig.~\ref{fig:ogw}.  
The label PH denotes  preheating after inflation \cite{tkach}, and the label
BH denotes a possible background from accummultaed black hole collapses\cite{ferrari}.
It is clear from Fig.\ref{fig:ogw} that the string cosmology background
can have a  much higher amplitude than all the other possible astrophysical
and cosmological sources of GW.

\section{Astrophysical and cosmological bounds}\label{section:second}

This section follows closely the discussion in \cite{moriond}. At the
moment, the most restrictive observational constraint on the spectral parameters
comes  from  the standard model of big-bang nucleosynthesis (NS) \cite{ns1}. 
This restricts the total energy density in gravitons to less than that of
approximately one massless degree of freedom in thermal equilibrium. This bound
implies that \cite{ba} \be
\int \Omega_{\rm GW}(f) d \ln f  = \Omega_{\rm GW}^{\rm S} \left[
\frac{1}{3}+ \frac{1}{\beta}\left( \left(f_1/f_{\rm S}
\right)^\beta-1\right)\right] < 0.7 \times 10^{-5} {\rm h}^{-2}_{100}.
\label{nucleo}
\ee
where we have assumed an allowed $N_\nu=4$ at NS, and have substituted
in the spectrum (\ref{e:approx}). The NS bound and additional cosmological 
and astrophysical bounds are shown in Fig.~\ref{fig:cabounds}, where $h_{100}$
was set to unity.

The line marked ``Quasar" in Fig.~\ref{fig:cabounds} corresponds to a bound 
coming from quasar proper motions. A stochastic background of gravity waves 
makes the signal from distant quasars scatter randomly on its way to earth.
This may cause quasar proper motions. An upper bound on quasar proper
motions can be translated into an upper bound on a stochastic
background~\cite{gwinn}. A typical strain $h$ may induce  proper motion $\mu$,
$h/f\sim \mu$. The sensitivity reached was approximately micro arcsecond per
year \cite{gwinn},  corresponding to a dimensionless strain of 
about $h\sim 5 \times 10^{-9}$
at frequencies below the observation time: approximately (20 years)$^{-1}\sim
5\times 10^{-9}Hz$, leading to $\Omega_{GW}\laq 0.1 h_{100}^{-2}$. Future
improvement in astrometric measurements could improve this bound substantially
\cite{gaia}. 

The line marked ``COBE" in Fig.~\ref{fig:cabounds} corresponds to the 
bound coming from energy density
fluctuations in the cosmic microwave background,  which can be
expressed in terms of the measured temperature fluctuations $\Delta T/T$, and
the fractional energy density in photons $\Omega_\gamma$ 
$\Omega({\rm perturbations})\simeq 
(\frac{\Delta T}{T})^2  \Omega_\gamma\sim 10^{-10}\times  10^{-4}=10^{-14}
h_{100}^{-2}$. Since it is known \cite{turner} that $\Omega_{GW}\laq 0.1 
\Omega({\rm perturbations})$,  it follows that $\Omega_{GW} h_{100}^2 \laq
10^{-15}$ at frequencies  $10^{-18} h_{100} Hz - 10^{-16}h_{100} Hz$.

The curve marked ``Pulsar" represents the bound coming from millisecond pulsar 
timing~\cite{td}.  Assuming known distance and signal emission times, the pulsar functions
as a giant one-arm interferometer. The statistics of pulse arrival time residuals
$\Delta T$, puts an upper bound on any kind of noise in the
system, including a  stochastic background of GW. The typical strain
sensitivity is  $h\sim \frac {\Delta T}{T}$, where $T$ is the total observation
time, reaching by now 20 years $\sim 6 \times 10^{8} sec$ and $\Delta T\sim 10
\mu s$  is the accuracy in measuring time residuals. Translated into
$\Omega_{GW}(f)$, this yields the bound shown in the figure, which is most restrictive
at frequencies $f\sim 1/T \sim 5\times 10^{-9} Hz$.
 
\begin{figure}
\begin{flushleft}
\hspace{-.2in}
\psfig{figure=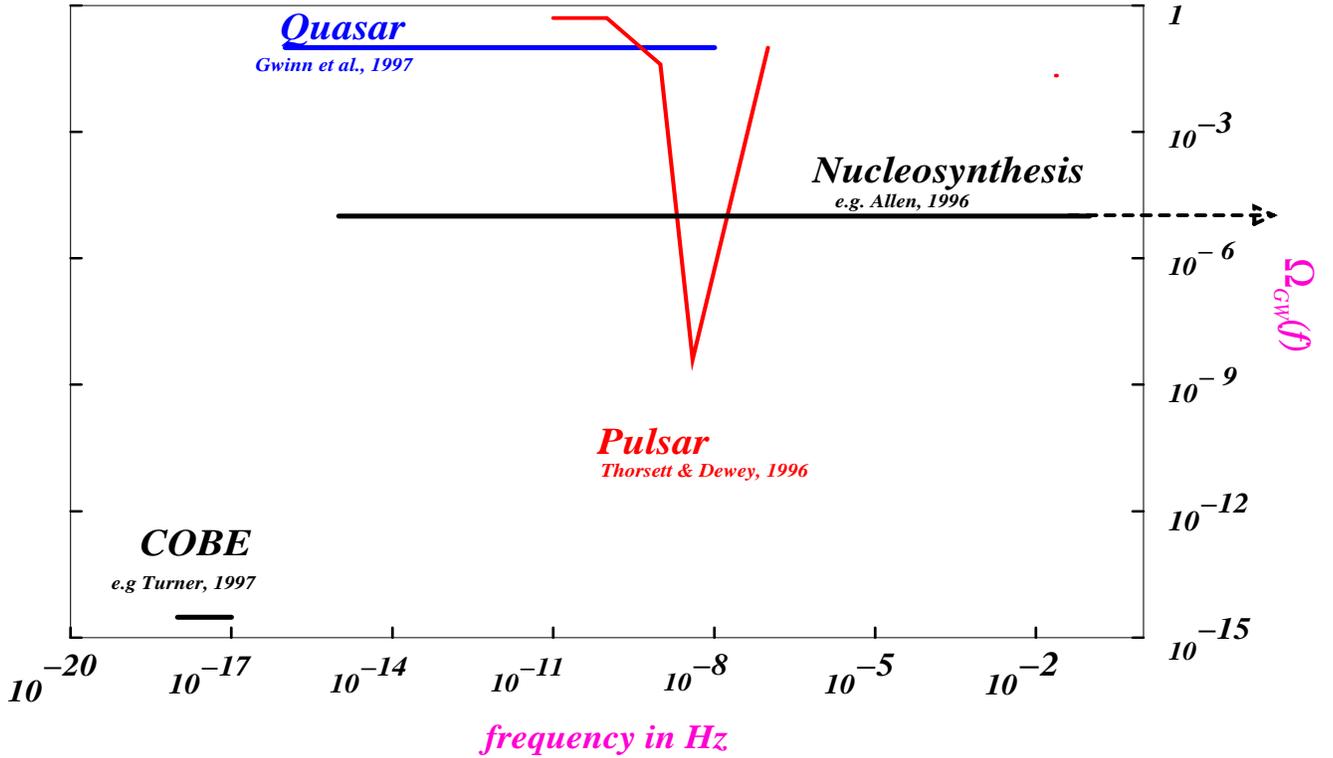,width=5.2in,height=6.5in}
\end{flushleft}
\vspace{-2.1in}
\caption{Cosmological and astrophysical upper bounds on the cosmic
gravitational wave background.
\label{fig:cabounds}}
\end{figure}

Notice that all the existing bounds, except from the NS bound which bounds the
total energy density, are in the very low frequency range, while
the expected signal from string cosmology is in a higher frequency range. The
bounds are therefore not very restrictive.

\section{Detecting a string cosmology stochastic gravitational wave
background}\label{section:third}

The principles of using a network of two or more gravitational wave antennae to
detect a stochastic background of gravitational radiation are by now well known
\cite{allen,stoch,vitale}.   The basic idea is
to correlate the signals from separated detectors, and to search for a
correlated strain produced by the gravitational wave background, which
is buried in the instrumental noise.  After correlating signals for time 
$T$ the ratio of signal
to noise   is given by 
\be \label{e:sovern}
\left( {S \over N} \right)^2 =
{9 H_0^4 \over 50 \pi^4} T \int_0^\infty df \>
{\gamma^2 (f) \Omega_{\rm GW}^2(f) \over f^6 P_1(f) P_2(f)}.
\ee  
The instrument noise in the detectors is described by the one-sided
noise power spectral densities, in units of $1/Hz$, $P_i(f)$. 
The dimensionless overlap reduction function $\gamma(f)$ is determined by the
relative locations and orientations of the two detectors \cite{allen}.  

It is useful to consider the approximation in which $P_1(f)$, $P_2(f)$ have a
maximum sensitivity at a common frequency $f_{ms}$ and a bandwidth $\Delta f$. 
Then 
\be \label{e:asovern} \left( {S \over N}
\right)^2 = {9 H_0^4 \over 50 \pi^4} T \Delta f 
{\gamma^2 (f_{ms}) \Omega_{\rm GW}^2(f_{ms}) \over f_{ms}^6 
P_1(f_{ms}) P_2(f_{ms})}.
\ee  
An obvious
remark is that both detectors need to have an overlapping frequency range around
their maximum sensitivity frequency, otherwise it is impossible to perform a
meaningful correlation experiment. To ensure a common frequency range, some
amount of tunning flexibility is very important.

We would like to highlight a few specific points about the string cosmology
background. In my opinion, one should look for the dilaton-driven signal even
though the string phase signal could be higher. The dilaton-driven
spectrum has the advantage that the spectrum is theoretically clean, and
therefore, if the $f^3$ dependence of the spectrum could be established it can
provide a clean experimental signal for detection.
Looking specifically at the sensitivity for detecting the spectrum produced 
during the dilaton-driven phase we obtain
\be \label{e:sasovern}
\left( {S \over N} \right)^2 =
{9 H_0^4 \over 50 \pi^4} T \Delta f
{\gamma^2 (f_{ms})\  {\Omega^S_{\rm GW}}^2 \over f_S^6 P_1(f_{ms})
P_2(f_{ms})}, 
\ee  
provided $f_{ms}<f_S$.

From equation (\ref{e:sasovern}) we can draw the following lessons.
An obvious conclusion is that it pays to increase the observation
time $T$. For a given $S/N$ the reach in $\Omega$ increases as $\sqrt{T}$.
Another observation is that it pays to increase sensitivity even if it comes at
the expense of bandwidth \cite{pizzella}. This is because the 
signal to noise ratio goes up linearly with the maximal sensitivity
of each detector $P_1(f_{ms})$, $P_2(f_{ms})$ but only increases as the square
root of the bandwidth. A conclusion that is perhaps not obvious is that it is
better  to search at the highest frequency, if the same sensitivity in $1/Hz$
can  be obtained. This is because the background from astrophysical 
sources is smaller at higher frequencies, so a detection at higher frequency 
provides  a cleaner signal.
Finally,  it is helpful to have as many detectors as near by as possible, 
without introducing correlated noise. 
Additional pairs of detectors do not add sensitivity because
the background is Gaussian \cite{allenrom}, therefore there is no additional
information in higher-point correlation functions. They do however increase the
level of confidence in the case of detection and provide a good way of reducing
local sources of noise. Tunable detectors could provide an opportunity to
verify the spectral shape and are therefore essential.

\subsection{Large detectors}

The LIGO project is building
two identical detectors,  the ``initial"
detectors. These detectors will be upgraded to so-called ``advanced" detectors. 
 Since the two detectors are identical in design, $P_1(f)=P_2(f)$.  
The design goals for the detectors specify these functions \cite{science92}. 
The design  noise power spectrum for the Virgo detector \cite{virgo} and 
of other large interferometers, GEO 600 \cite{geo600}
and TAMA 300 \cite{tama300} and the noise power spectral
densities of operating  and planned resonant mass GW detectors (``bars")
\cite{naut,auriga} are also known.
The overlap reduction function $\gamma(f)$  is
identical for both the initial and advanced LIGO detectors, and has been
determined for many pairs of GW detectors \cite{allen,vitale}. 

 \begin{figure}
\begin{flushleft}
\hspace{-.5in}
\psfig{figure=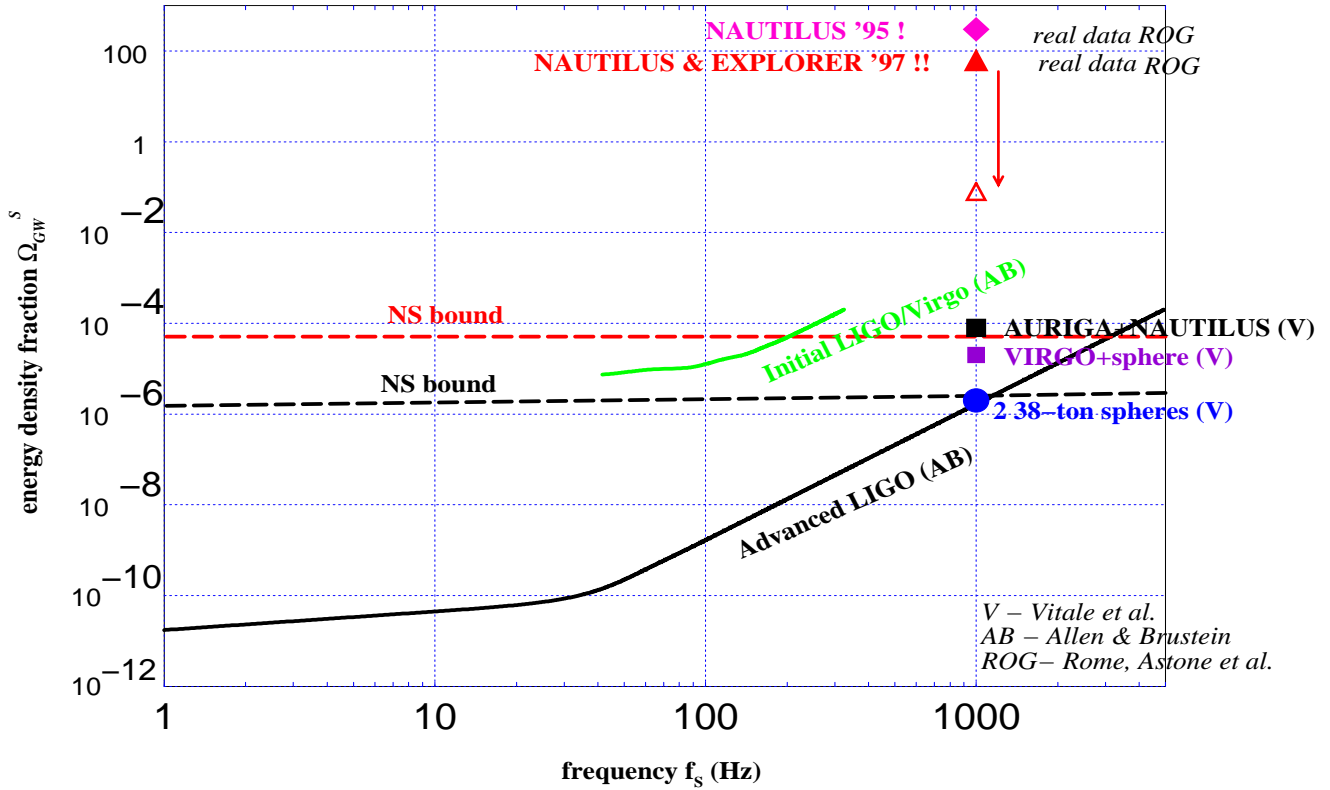,width=5.2in,height=6.5in}
\end{flushleft}
\vspace{-2.3in}
\caption{Detection sensitivity of relic GW by operating and planned GW 
detectors. The interesting region of parameter space is below
the ``NS bound" lines.\hfil\hfil\hfil
\label{fig:gensens}}
\end{figure}

Making use of the prediction from string cosmology (\ref{e:approx}), 
we may use equation (\ref{e:sovern}) to assess the detectability of this stochastic
background.  For any given set of parameters we may numerically
evaluate the signal to noise ratio $S/N$; if this value is greater than
$1.65$ then with at least 90\% confidence, the background can be
detected by a given pair of detectors.  The regions of detectability in parameter
space are shown in Fig.~\ref{fig:gensens}. The region below the  NS bound lines
and above the advanced LIGO curve is the region of interest. Two NS bounds are shown, 
the upper, more relaxed bound, assumes no GW production during the 
string phase \cite{ba}. The points at 1 KHz come from operating
and planned resonant mass detectors.
Some are taken from real experiments, an upper bound from a single detector run
 \cite{nautstoch}, and the first modern 12.5 hours correlation experiment
 between Nautilus and Explorer \cite{nautexplo97}. The arrow points
to a hollow triangle showing by how much the correlation 
experiment can be improved if Nautilus works properly and the experiment could 
be done for one year. Other points are from theoretical
calculations~\cite{vitale}. 
 For Fig.~3 we have assumed ${\rm h}_{100}=0.65$
and $H_{*}=5 \times 10^{17} \> {\rm GeV}$. 

\subsection{Small detectors}

The expected GW signal from string cosmology could have  substantial
power at high-frequencies. It is therefore tempting to explore the possibility
to detect it with small devices, which could be more sensitive at high
frequencies. The idea, in principle, is very simple. Build the most accurate
long-lived two-level system possible. A practical way to do this is to take two
identical  resonators with the highest finesse, and couple them weakly.
The weak coupling splits the resonance level  into two near by levels. Then load
the lower ``pump" level with as many target particles,  photons, atoms etc. (they
have to be bosons, of course) and wait for a gravity wave to come along and
knock one of the particles up to the upper level. The resonators can be
electromagnetic \cite{gris}, in the microwave or optical bandwidth, or perhaps a
coherent atomic system. 

\subsubsection{Microwave cavity detectors}

A practical design of a two-level system is achieved by taking two
superconducting microwave cavities and coupling them weakly \cite{mwcav}. The
lower symmetric level is the pump level, and the upper antisymmetric level is
the ``output" level. 
 For a monochromatic GW  of frequency
$f_{GW}$, the sensitivity is the  following \cite{mwcav},
\begin{equation}
h(f_{GW})=\frac{\Delta \ell}{\ell}\sim 2\left(\frac{U_{\hbox{\it a}}}
 {U_{\hbox{\it s}}}\right)^{1/2} \frac{1}{Q} \left[ 1+4Q^2
\left(\frac{\Delta-f_{GW}}{f_{\hbox{\it a}}}\right)^2\right]^{1/2}
\label{mcsens}
\end{equation} 
 where $\Delta=f_{\hbox{\it a}}-f_{\hbox{\it s}}$. 
$f_{\hbox{\it a}}$, $f_{\hbox{\it s}}$ and
$U_{\hbox{\it a}}, U_{\hbox{\it s}}$ are the frequencies and energies stored in
the antisymmetric and symmetric levels respectively. For a burst of duration
$1/f_{GW}$ smaller than the lifetime of the resonance $Q/f_{\hbox{\it a}} $ the
sensitivity in eq.(\ref{mcsens}) becomes 
\begin{equation} h(f_{GW}) \sim 
2\left(\frac{U_{\hbox{\it a}}}{U_{\hbox{\it s}}}\right)^{1/2}  \frac{1}{Q}\left[
2 Q \frac{f_{GW}}{f_{\hbox{\it a}} }\right]= 4\left(\frac{U_{\hbox{\it
a}}}{U_{\hbox{\it s}}}\right)^{1/2} \frac{f_{GW}}{f_{\hbox{\it a}} } 
\end{equation} 
during the lifetime $Q/f_{\hbox{\it a}}$ of the resonance 
there are $N=(Q/f_{\hbox{\it a}}) f_{GW}$ such independent short bursts
and the sensitivity increases by a factor $\frac{\sqrt{ N}}{N}=
\left(f_{\hbox{\it a}}/f_{GW} Q\right)^{1/2}$. 
The sensitivity estimate now becomes
\begin{eqnarray}
\frac{\Delta \ell}{\ell}&\sim& 
4\left(\frac{U_{\hbox{\it a}}}{U_{\hbox{\it s}}}\right)^{1/2}
\frac{f_{GW}}{f_{\hbox{\it a}} } 
\left(f_{\hbox{\it a}}/f_{GW} Q\right)^{1/2}\nonumber \cr \\
&=& 4\left(\frac{U_{\hbox{\it a}}}{U_{\hbox{\it s}}}\right)^{1/2}
\frac{1}{\sqrt{Q}}\left(\frac{f_{GW}}{f_{\hbox{\it a}} } 
\right)^{1/2} 
\end{eqnarray} 
which, very optimistically,  at $f_{GW}\sim 10 KHz$, with a $Q\sim 10^{12}$, 
$U_{\hbox{\it a}}\sim 10^{-22}$  Watt, $U_{\hbox{\it s}}\sim 100$ Watt, 
$f_{\hbox{\it a}}\sim 10 GHz$ will give $h\sim
10^{-21}$, yielding a respectable sensitivity of  
$\Omega_{GW}\sim 10^{-4}h_{100}^{-2}$ for a correlation experiment lasting
one year, assuming a bandwidth of about $10 KHz$ without considering the thermal 
noise and selectivity criterion. Thermal noise, in particular, 
can be the real killer for such a detector.

A similar result is obtained by integration of the response function of the two
level system
 $\left[ 1+4Q^2 \left(\frac{\Delta-f_{GW}}
{f_{\hbox{\it a}}}\right)^2\right]$ against the density of gravitational
energy $\rho(f)\sim f^2 h^2(f)$ to estimate $U_{\hbox{\it a}}$
\cite{private}.

A prototype is being built presently, aiming to prove  the feasibility of
building a real GW detector of this type, and to verify that sensitivity
estimates are indeed reasonable \cite{paco}.

\subsubsection{Motion of free masses: the memory effect}

A massive object, initially at rest, will start to move in a random motion
under the influence of a stochastic background of GW, performing a sort of 
Brownian motion. Monitoring the position of the object over a length of time 
can therefore be used to detect the existence of a GW background. A possible 
setup can consist of two masses that are free to move, and a device that 
measures their relative distance, similar to an interferometer, with the 
important difference that masses are allowed to move freely.
The idea was discussed by Braginskii and Grishchuk \cite{brag}. 
Here we use a different method of obtaining the estimated sensitivity.

We would like to compute $\Delta \ell_{r.m.s.}=
\left(\left\langle \Delta \ell^2 \right\rangle\right)^{1/2}$, the average taken
over random realizations, or equivalently a time average. To evaluate
$\Delta \ell_{r.m.s.}$ we use the relation
$ h_{rms}^2=
\left\langle \frac{\Delta \ell^2}{\ell^2} \right\rangle$ and evaluate
$h_{r.m.s.}$,
\begin{equation}
h_{rms}^2=\frac{1}{2} \int_{-\infty}^\infty d f S_h (f) 
= \frac{3 H_0^2}{4 \pi^2}\int_0^\infty
df \frac{\Omega_{GW}(f)}{f^3}.
\end{equation}
Using the explicit form (\ref{e:approx}) of the spectral density, taking into
account only the contribution from the dilaton-driven phase we obtain
\begin{equation}
h_{rms}^2=\frac{3 H_0^2}{4 \pi^2 f_S^2} \Omega_{GW}^S.
\end{equation}
Plugging in some reasonable numerical values we obtain
\begin{equation}
h_{rms}^2 \simeq 10^{-54} (1 MHz/f_S)^2 (\Omega_{GW}^S/ 5 \times
10^{-7} h_{100}^{-2}), 
\end{equation}
leading to motions $\Delta \ell \sim 10^{-25} cm$ for $\ell\sim 100 cm$. The
surprise/disappointment is that the average displacement does not grow with time
as in ordinary Brownian motion. 

The same result can be obtained by computing the force exerted by the
stochastic background on the massive object using Newtonian mechanics,
$\vec F = m \vec a$. The force is given by the geodesic equation,
\begin{equation}
\frac{d^2x^j}{dt^2}_{waves}=-{R_{j0i0}}_{waves}x^i
\end{equation}
where
$R_{j0i0}=-\frac{1}{2} h_{ji}^{TT},_{00}$.

Looking for simplicity at the motion caused by a single TT component in one
direction, $\ddot{\vec\ell}\sim \ell \ddot{h}$, and the equation can be
integrated 
\begin{equation}
\frac{\Delta x^2}{x^2} =\langle \int^t \int {\ddot h}_{ij}(x,s) 
\int^t \int {\ddot h}_{ij}(x,s')\rangle
\end{equation}
leading up to some geometry factors to the same answer, in particular, showing
that the r.m.s. displacement does not grow with time.

This approach does not look too promising, but perhaps could be improved.

Our sensitivity analysis of detectors clearly points in favor of large
detectors,  at least at this point of technological development. However,
using small detectors is still a very interesting enterprise, because a discovery
of high frequency GW is a unique cosmological signature of a  high curvature
universe \cite{grishchuk,maggiore}, and could allow the cleanest detection of
the string cosmology background.

\hspace{.5in}
\section*{Acknowledgments}

I would like to thank my collaborators Bruce Allen, Maurizio Gasperini and 
Gabriele Veneziano. I would like to thank Vladimir Braginskii for drawing my
attention to the possible use of the ``memory effect". Thanks to many GW
experimentalists (too many to name everyone, sorry) for their help, and special
thanks to the  ROG collaboration and Pia Astone for access to their data.  This
work is supported in part by the  Israel Science Foundation administered by the
Israel Academy of Sciences and Humanities.

\end{document}